\documentclass[iop]{emulateapj}

\shorttitle{Death of Magnetic Features}
\shortauthors{Lamb et~al.}

\defcitealias{DeForest2007}{SMT-1}
\defcitealias{Lamb2008}{SMT-2}
\defcitealias{Lamb2010}{SMT-3}

\begin{document}

\title{Solar Magnetic Tracking. IV. The Death of Magnetic Features}
\author{D.~A. Lamb, T.~A. Howard, and C.~E. DeForest}

\affil{Southwest Research Institute, 1050 Walnut Street, Suite 300, Boulder,
CO, 80302, USA}

\email{derek@boulder.swri.edu}

\author{C.~E. Parnell}

\affil{School of Mathematics and Statistics, University of St.~Andrews,
St.~Andrews, Scotland, KY16~9SS, UK.}

\and

\author{B.~T. Welsch}

\affil{University of California, Berkeley, Space Sciences Laboratory, 7
Gauss Way, Berkeley, CA, 94720, USA.}

\begin{abstract}
The removal of magnetic flux from the quiet-sun photosphere is
important for maintaining the statistical steady-state of the magnetic
field there, for determining the magnetic flux budget of the Sun, and
for estimating the rate of energy injected into the upper solar
atmosphere. Magnetic feature death is a measurable proxy for the
removal of detectable flux, either by cancellation (submerging or
rising loops, or reconnection in the photosphere) or by dispersal of
flux. We used the SWAMIS feature tracking code to understand how
nearly $2\times10^{4}$ detected magnetic features die in an hour-long
sequence of \emph{Hinode}/SOT/NFI magnetograms of a region of quiet
Sun. Of the feature deaths that remove \emph{visible} magnetic flux
from the photosphere, the vast majority do so by a process that merely
disperses the previously-detected flux so that it is too small and too
weak to be detected, rather than completely eliminating it. The
behavior of the ensemble average of these dispersals is not consistent
with a model of simple planar diffusion, suggesting that the dispersal
is constrained by the evolving photospheric velocity field. We
introduce the concept of the \emph{partial lifetime} of magnetic
features, and show that the partial lifetime due to Cancellation of
magnetic flux, 22~h, is 3 times slower than previous measurements of
the flux turnover time. This indicates that prior feature-based
estimates of the flux replacement time may be too short, in contrast
with the tendency for this quantity to decrease as resolution and
instrumentation have improved. This suggests that dispersal of flux to
smaller scales is more important for the replacement of magnetic
fields in the quiet Sun than observed bipolar cancellation. We
conclude that processes on spatial scales smaller than those visible
to \emph{Hinode} dominate the processes of flux emergence and
cancellation, and therefore also the quantity of magnetic flux that
threads the photosphere.
\end{abstract}

\keywords{Sun: granulation --- Sun: photosphere --- Sun: magnetism}

\label{sec:Introduction}
\section{Introduction}

The solar photosphere contains a patchwork of magnetic field regions,
whose size varies from sunspots, sometimes visible to the naked eye
from Earth, to below the spatial resolution limit of current
telescopes \citep[e.g.][]{Schrijver2000}.  While sunspots are located
on the solar disk in a predictable pattern throughout the solar cycle,
the smaller magnetic regions are roughly evenly distributed across the
Sun at all times \citep{Harvey1993PhDT,Hagenaar2001}, and are
constantly in motion.

Measurement of the behavior of small magnetic features on the
photosphere is limited, partly by the spatial and temporal resolution
of the observing instruments, and partly by the difficulty of
following visual features that do not behave exactly like discrete
physical objects.  Tracking these features was first performed by the
human eye \citep[e.g.][]{HarveyHarvey1973}, and some groups still use
that method \citep[e.g.][]{Zhou2010} despite the subjectivity and
potential unknown bias of human interpretation.  Experience has shown
\citep{DeForest2007} that even automated methods of solar feature
tracking, produced by different authors with the intention of
reproducing others' results, have myriad built-in assumptions and
subjectivity of their own unless great care is taken in specifying the
algorithm exactly.  This result is not limited to the tracking of
magnetic features in solar magnetograms
\citep{Welsch2007shootout,DeRosa2009}.

In the first part of the present series on solar magnetic tracking
\citep[SMT-1,][]{DeForest2007}, we described four different magnetic
feature tracking algorithms, showed how small differences between the
algorithms affect derived physical quantities such as the flux and
lifetime distribution of magnetic features, and recommended a standard
methodology for their tracking.  In the second
\citep[SMT-2,][]{Lamb2008}, we used the SWAMIS code to track features
in a series of \emph{SOHO}/MDI \citep{Scherrer1995} high-resolution
magnetograms. We found that the vast majority of newly-detected flux
was due to the coalescence of previously existing magnetic field,
rather than bipolar flux emergence from the solar interior. Those
results agreed with those of \citet{Muller2000} and were confirmed in
the third part of this series \citep[SMT-3,][]{Lamb2010}, which compared
MDI data with simultaneous higher-resolution magnetogram data from
\emph{Hinode}/SOT/NFI \citep{Kosugi2007,Tsuneta2008}.  In
\citetalias{Lamb2010} we also showed, through a similar analysis to
that in \citetalias{Lamb2008}, that \emph{Hinode} does not resolve the
fundamental scale of flux emergence: we found that most new magnetic
features, both by number and by entrained magnetic flux, arise through
coalescence of unresolved magnetic flux into larger concentrations
that can be resolved by the instrument.  This result is in agreement
with theoretical results of prior authors \citep[e.g.][\&\ references
  therein]{Schrijver1997,Simon1995,Simon2001} who explored cross-scale
processes and their role in sustaining the Sun's magnetic network. It also
highlights work showing that that the smallest observable features
dominate the magnetic flux balance at all currently observable scales
\citep{Parnell2009}, and that much of the solar magnetic flux is as
yet undetectable \citep{Krivova2004,TrujilloBueno2004}.

Since the quiet-sun photospheric magnetic field exists in an
statistical steady state, understanding the process by which magnetic
flux is removed from the photosphere is just as important as
understanding the process by which it is introduced. The death of
visible magnetic features is the best available proxy for this
process. To understand the processes by which magnetic flux is removed
from the photosphere, we have re-analyzed the same
\emph{Hinode}/SOT/NFI dataset that was used for \citetalias{Lamb2010},
examining the relationship between feature birth and death, and the
principal processes by which features die.

The quiet sun photospheric flux budget can be characterized roughly by
just two quantities: the total unsigned magnetic flux threading the
photosphere, and the rate at which flux is introduced or removed. The
ratio of the two quantities yields a ``replacement time''---a
characteristic timescale over which all of the quiet sun photospheric
flux will be replaced with new flux. But even this seemingly simple
calculation is difficult in practice, because the two elements of the
quotient are both hard to measure. Resolution effects in Zeeman-effect
line-of-sight magnetograms drastically reduce the estimated total flux
threading the photosphere, because the necessary averaging over each
pixel is a \emph{signed} average (\citealp[e.g.][\&\ references
  therein]{Harvey1993PhDT};
\citealp[also][]{PietarilaGraham2009}). This has led to a general
increasing trend in estimates of the total unsigned flux as
instruments improve. Hanle effect measurements
\citep{TrujilloBueno2004} are not subject to the averaging problem but
involve assumptions of their own \citep{PietarilaGraham2009}. These
resolution effects also influence the amount of flux deemed to have
been introduced or removed from the photosphere.

Turnover time estimates have typically been made by measuring feature
lifetimes -- either visually \citep[e.g.][]{HarveyHarvey1973,Zhou2010}
or algorithmically
\citep[e.g.][]{Hagenaar1999,Hagenaar2008,Hagenaar2009,Iida2012}.  But
feature lifetimes do not necessarily correspond well to actual
creation and destruction of the flux that the features contain.  In
particular, the visual process of Appearance dominates the
distribution of magnetic features in the photosphere
\citepalias{Lamb2008} but is caused by rearrangement (coalescence) of
existing, previously unresolved magnetic flux into concentrations
sufficiently large to be resolved \citepalias{Lamb2010}.  Similarly,
it is possible for features to Disappear by dispersal (the opposite of
coalescence), which eliminates visually measurable magnetic flux but
does not itself alter the total number of field lines threading the
photosphere. This effect means that na\"{\i}ve feature-based estimates
of the flux replacement time may be too short by up to an order of
magnitude.

Further, the process by which flux is actually removed from the
photosphere is important because it drives several processes important
for coronal heating and structure (e.g.
\citealt{Longcope1999,Parker1988,LopezFuentes2006};
\citetalias{DeForest2007}).  Thus, understanding the primary scale (or
scale distribution) on which flux removal takes place is important
to understanding the energy release mechanisms and magnetic structure
that give rise to the corona and shape it.

\label{sub:IntroRemoval}
\subsection{Removal of Magnetic Flux from the Photosphere}

Magnetic flux is conserved, so only two processes can reduce the
signed flux threading a particular patch of photosphere: annihilation,
in which a collection of opposing flux enters the patch; or dispersal,
in which a collection of like-signed flux leaves it. These processes
are reflected in similar events that affect visible magnetic features,
and are defined in \citetalias{DeForest2007}.  As in our previous
work, we capitalize the names of feature birth and death \emph{events}
to emphasize that they represent observables (and are localized in
time), whereas true physical \emph{processes} are lowercase.  For
example, Cancellation involves two visible opposing features
converging and shrinking as they interact
\citep[e.g.][]{Livi1985,Wang1988,Priest1994}, Fragmentation
involves a single feature breaking up into multiple smaller ones by
dispersal, and Disappearance may involve dispersal into undetectably
small features \citepalias{DeForest2007}.

Cancellation observed in magnetograms may be the manifestation of one
of three physical processes: 1) the submergence of $\Omega$-shaped
loops into the solar interior; 2) the rise of U-shaped loops into the
upper solar atmosphere; 3) magnetic reconnection occurring in the
``magnetogram layer'' of the solar atmosphere itself.  Reconciling
these, even along the often-studied magnetic neutral lines of solar
active regions, is difficult due to a lack of absolute velocity
measurements in the photosphere \citep{Welsch2013}.

The dispersal of flux arises from advection of magnetic flux by the
turbulent motion of the photosphere. \citep[e.g.][]{Leighton1964}. It
leads to a random walk of individual magnetic field lines over the
surface of the Sun, which leads to diffusion-like processes, although
the characteristics and importance of these processes have remained a
matter of discussion for over 40 years and are unlike normal diffusion
\citep[e.g.][]{Smithson1973,Simon1995,Hathaway1996,Berger1998,Hagenaar1999,Cadavid1999,Parnell2001,Abramenko2011}.

The removal of visible magnetic features can be a result of both
cancellation and dispersal of flux, but these processes differ in
important ways. Of the two processes, cancellation is more likely to
release energy into the solar atmosphere by driving reconnection, and
is the only process that can truly remove magnetic flux from the
surface of the Sun. The two processes also contribute to very
different statistical behavior of the overall magnetic field of the
Sun. Magnetic field dispersal is often treated as a diffusive process
in which evolution of the field has been presumed to approximate a
diffusion law $dB_r/dt \propto \nabla^2B_r$ \citep{Leighton1964}. In
contrast, cancellation causes removal of flux from the photosphere, is
also part of the emergence/cancellation quasi-diffusion process that
forms the fine-scale field, and may constitute the small-scale
dynamo. Quasi-diffusion follows a different functional form than
diffusion. Its behavior depends on the statistics of emergence and
degree of mixing between signs in the existing field
\citep[e.g.][]{Schrijver1997,Simon2001,Abramenko2011}.

In the present paper, we perform an analysis of feature death (as
defined by us in \citetalias{DeForest2007}) in a sequence of
magnetograms taken with the \emph{Hinode} Narrowband Filter Imager
(NFI) instrument \citep{Tsuneta2008}.  In Section \ref{sec:Data
  Processing}, we discuss the data and the processing steps we used;
in Section \ref{sec:Results}, we show results including the
distribution of feature death types by number and by flux, and that
the time evolution of an ensemble of Disappearing features does not
follow the familiar planar diffusion equation; and in Section
\ref{sec:Discussion} we discuss these results and their implications for
the solar dynamo.

\section{Data Processing and Selection Criteria}
\label{sec:Data Processing}
\subsection{Data}
\label{sub:Data}
Details of the selection and preparation of the dataset are provided
in \citetalias{Lamb2010}.  The data used here are exactly the same as
the short-duration NFI dataset in that work. In brief, the data are an
hour-long sequence of \emph{Hinode}/SOT/NFI Na I D 5896~\AA\ line-wing
magnetograms from 2007-06-24, 22:09UT - 22:08. The images have a
cadence of 1~minute, and a pixel scale of $0.16''$. The Stokes V/I
images were calibrated to a simultaneous high-resolution
($0.6''~\textrm{pixel}^{-1}$) MDI magnetogram using a factor of
$6555~G$. Our factor is smaller than the 16~kG found by
\citet{Zhou2010} and the 9~kG found by \citet{Iida2012}.  To compare
our results to those of either set of authors thus requires
multiplying our reported values of the magnetic field strength or
magnetic flux by the ratio $\frac{9}{6.5}$ or $\frac{16}{6.5}$.  The
rest of the preprocessing included cosmic ray despiking, derotation
(including cropping), temporal and spatial smoothing, and an FFT
motion filter which further reduces noise by $\sim20\%$ and rejects
solar p-modes. See \citetalias{Lamb2010} for further details.

We analyzed the magnetograms using the SWAMIS feature tracking code.
Details of its function and comparison with other tools are provided
in \citetalias{DeForest2007}.  We used the 2012-Aug-29 version of
SWAMIS\footnote{available at
  \anchor{http://www.boulder.swri.edu/swamis}{http://www.boulder.swri.edu/swamis}}.
Since \citetalias{Lamb2010} we have improved the code in two important
ways, summarized below.

The first change is in the initial tracking step, discrimination
\citepalias[see][\S~2.2]{DeForest2007}, in which pixels in features
are separated from background noise. The dual-threshold hysteretic
discriminator in SWAMIS initially included only pixels that are higher
than the high threshold before searching for neighboring pixels above
the low threshold. SWAMIS now adds to this initial high-threshold list
the central pixel in any group of three or more adjacent pixels that
are all above the low threshold. The rationale behind this change is
that in an image of random Gaussian-distributed noise, and for a
sufficiently high low threshold (e.g., $3\sigma$), the probability of
three adjacent pixels being above the noise floor is practically
zero. In particular, in a test dataset of ten ($1000\times1000$)-pixel
images of pure noise, there were no groups of three or more adjacent
pixels of the same sign above $3\sigma$. This improvement allows the
code to detect persistent weak features that might never have a single
pixel higher than the high threshold.

The second change is in the feature identification step
\citepalias[see][\S~2.3]{DeForest2007}. We fixed a recently-introduced
error in SWAMIS' downhill discriminator that caused feature ID numbers to
sometimes not stop at a local minimum. Obviously this does not affect
any previous results that used the ``clumping'' discriminator
\citep{Parnell2009}, and it does not affect our previous work in this
series \citepalias{Lamb2008,Lamb2010} because that work was mainly
concerned with features that were not touching other features, and so
did not deal with local minima as boundaries between features. Other changes
since the 2008-May-19 version of SWAMIS are minor and mostly focus on
performance enhancements and usability improvements.

The parameters used in the feature tracking are also unchanged from
\citetalias{Lamb2010}, and are: detection thresholds of 18 \& 24~G,
the ``downhill'' method of feature identification and a per-frame
minimum size filter of 4~pixels. Per-feature filters included lifetime
$\ge$ 4 frames, largest size $\ge$ 4 pixels (which is redundant due to
the per-frame minimum size filter). These per-feature filters do not
apply to features that are spatially immediately adjacent to another
feature at any point during their life; this prevents the rejection of
features that are obviously part of a larger magnetic field
concentration. In the birth and death classification, we look for
pairs of features separated by at most 5 pixels, and require that the
changes in flux among interacting features agree to within a factor of
2 in order to approximately satisfy flux conservation.

\subsection{Event Identification}
\label{sub:EventIdentification}
A feature birth event occurs when a visual feature exists in a given
frame and can not be seen in the previous frame
\citepalias{Lamb2008}. Similarly, a feature death event occurs when a
visual feature exists in a given frame and can not be seen in the next
frame. We classify feature deaths with the same criteria as births,
reversed in time. \citetalias{Lamb2008}, continuing with established
terms describing the processes that dictate the behavior of magnetic
features \citep[e.g.][]{Parnell2001}, used five terms to categorize
types of describe the observation of feature birth: Appearance,
Emergence, Fragmentation, Complex, and Error. Likewise, we follow the
terms used by \citetalias{Lamb2008} for types of feature death:
\begin{itemize}
\item \textbf{Disappearance}, where a feature dies with no other features in
the vicinity; 
\item \textbf{Cancellation}, where a feature dies near another of opposite polarity,
and the flux is approximately conserved; 
\item \textbf{Merger}, where a feature dies near another another of the same
polarity and flux is approximately conserved; 
\item \textbf{Complex}, involving multiple features where one satisfies the Cancellation
criteria while another satisfies Merger; 
\item \textbf{Error}, where a dying feature satisfies the proximity and polarity
of a Cancellation or Merger but does not approximately conserve
flux;
\item \textbf{Survival}, where the last frame in which a feature is
  detected is the last frame of the dataset.
\end{itemize}

\begin{figure*}
\includegraphics[height=0.75\textheight]{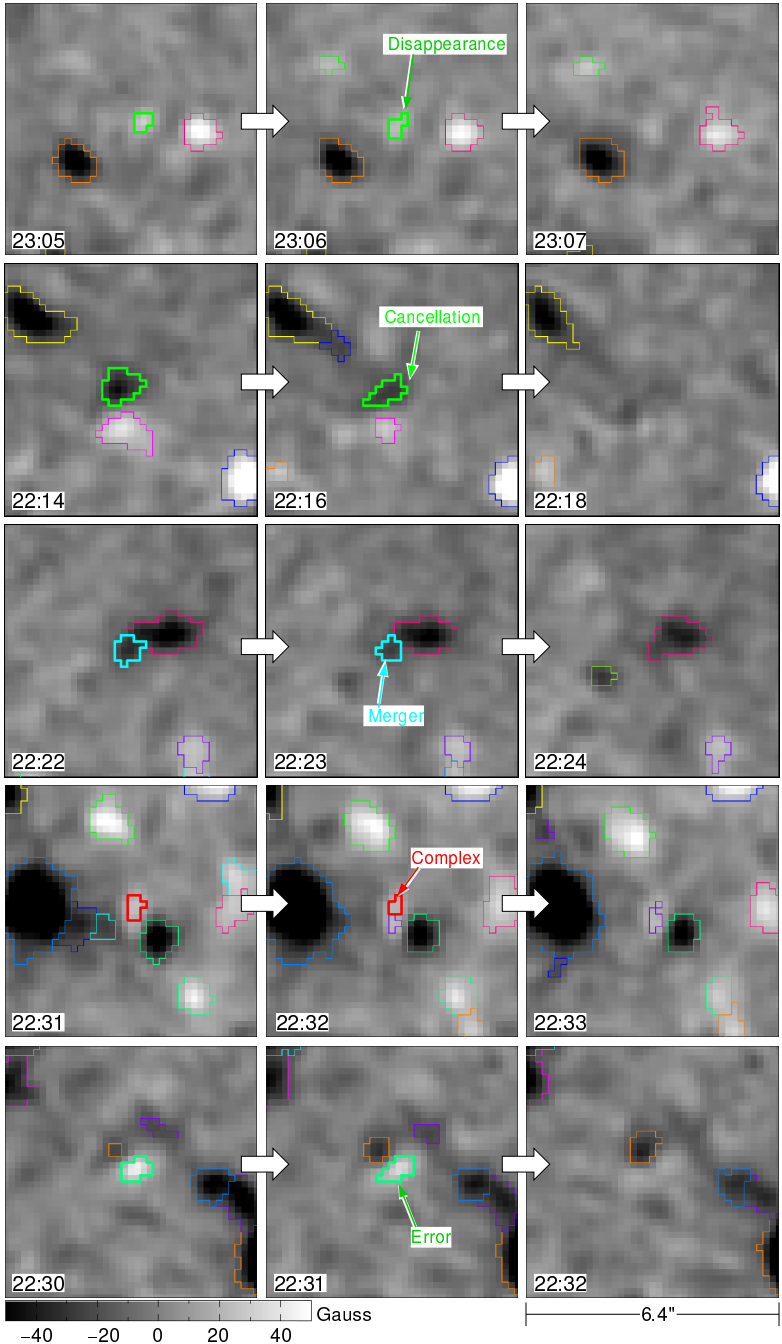}
\caption{Examples of \emph{Hinode}/SOT/NFI images of five death types
  studied in this paper.  Features are highlighted by the bordering
  color. Top Row: Disappearance, where a feature dies with no other
  features within 5~pixels; Second Row: Cancellation, where a feature
  dies near another of opposite polarity while approximately
  conserving flux; Third Row: Merger, where a feature dies near
  another of the same polarity while approximately conserving flux;
  Fourth Row: Complex, which involves multiple features where one
  satisfies the Cancellation criteria and another satisfies
  Merger; Bottom Row: Error, where a Cancellation would have
  been identified but the flux of the opposite-polarity feature
  \emph{increased} instead of decreased. Each image shows a
  $6.4''\times6.4''$ area on the photosphere and with the gray scale
  saturating at $\pm50~\textrm{G}$.}
\label{fig:death-examples} 
\end{figure*}

Figure~\ref{fig:death-examples} shows an example of each of
Disappearance, Cancellation, Merger, Complex and Error feature death
categories.

\section{Results}
\label{sec:Results}

\subsection{Summary of Detected Birth \& Death Events}
\label{sub:Event-Summary}
We identified 18297 features during the selected time period.  There
were 2088 features identified at the beginning of the dataset (birth
Survival) and 2729 features identified at the end of the dataset
(death Survival). There were 175 features that lived through the
entire dataset and thus were classified as both birth Survival and
death Survival.

Table~\ref{tab:N-stats} shows the percentage (of the total number of
features) for each birth / death type combination. Note that the 4
combinations of Fragmentation, birth Error, death Error, and Merger
account for 60\% of all features. We speculate that most of the birth
Error and death Error events are simply fragmentations \&
mergers\footnote{These are intentionally lowercase. Since the event
  categories are defined using strict criteria, similar events that do
  not meet those criteria are not capitalized.} for which flux was not
conserved in our simple two-feature interaction model.

\begin{figure*}
\includegraphics[width=1.0\textwidth]{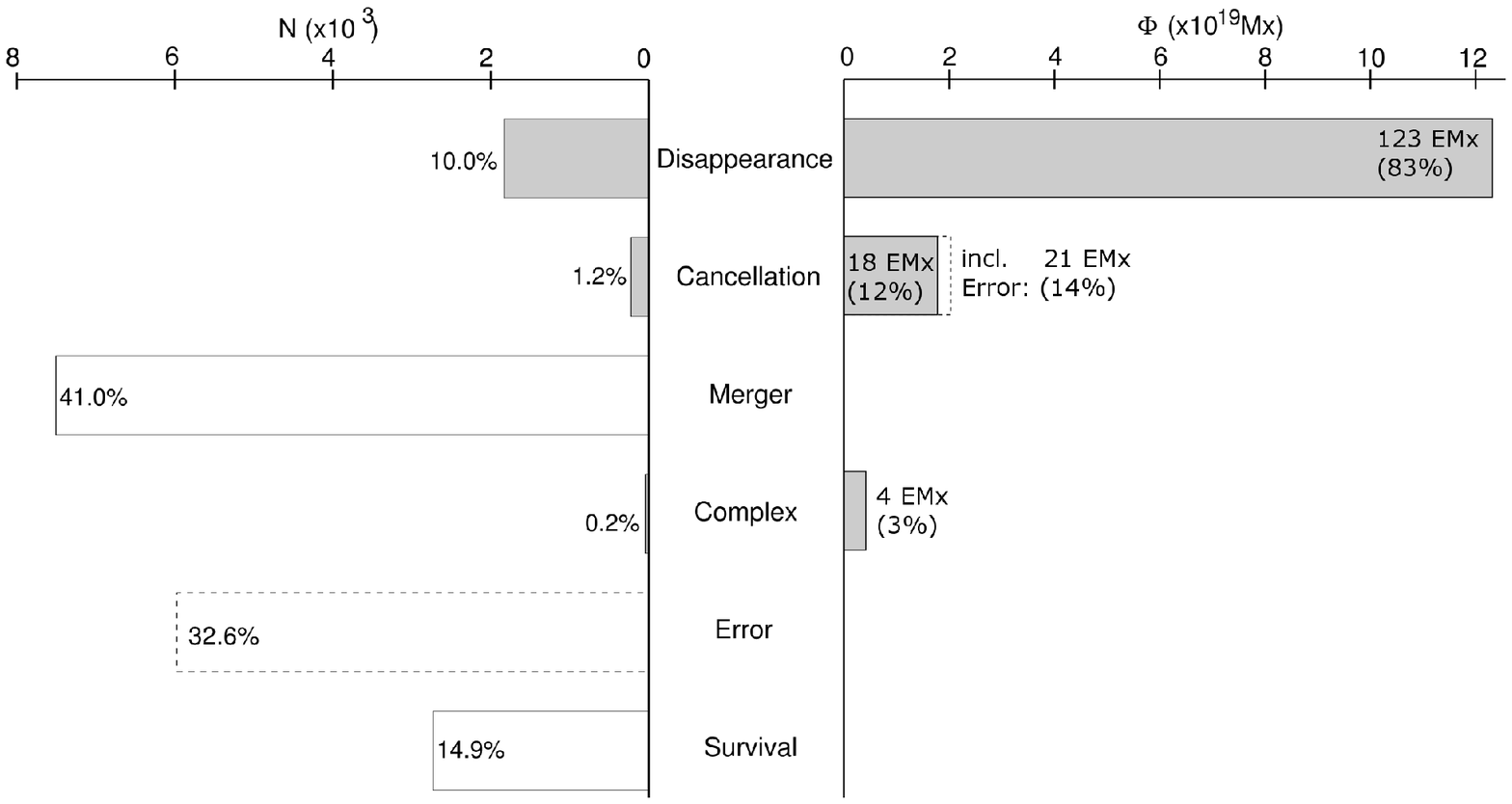}
\caption{Left: Histogram showing the number distribution of death type
  for the 18297 features detected using SWAMIS (i.e., bottom row in
  Table~\ref{tab:N-stats}). The gray shading shows those three event
  types that remove flux from the photosphere: Disappearance,
  Cancellation, and Complex. Right: Histogram showing the flux
  ($\Phi$) removed by each of the death processes. See
  \S~\ref{sub:Event-Summary} text.}
\label{fig:N-Number} 
\end{figure*}

Figure~\ref{fig:N-Number} shows the distribution of death type as a
histogram, and also the flux removed according to death type. We
accounted for flux by considering each feature to contain the maximum
flux achieved in any single frame throughout its lifetime.  The
Disappearance, Cancellation and Complex event types are shaded in
Figure~\ref{fig:N-Number}.  For the Error events, we estimate their
contribution to the Cancellation and Merger classes by distributing
them in the same proportion to that of the flux proportion of the
Cancellation and Merger events. The vast majority of this proportion
is from the Merger class and thus results in no flux removed. The
additional contribution to the Cancellation events due to this is
shown as the dashed extension. We disregard the Survival, Merger, and
the remainder of the Error death events for the following reasons:

\begin{itemize}
\item\textbf{Survival:} These features survive beyond the time frame
  of the dataset and therefore do not remove flux during the study;
\item\textbf{Merger:} These features remove no flux from the system
  since they lose their identity but not their flux by being absorbed
  into a like-polarity feature;
\item\textbf{Error:} In their SWAMIS-detected form, these are not
  physical events but either Cancellation or Merger events, although
  we lack sufficient information to identify which.
\end{itemize}

\begin{table*}
\tabletypesize{\scriptsize}
\begin{center}
\caption{Feature event history table: for each combination of birth \& death type, the percentage of all 18297 features that were born and died in that combination.}
\tabcolsep=0.11cm
\begin{tabular}{llllllll}
\tableline\tableline 
Birth \textbackslash{} Death Type & \textbf{Disappearance} & Cancellation & Merger & Complex & Error & Survival & Total \\
\tableline\tableline 
Appearance    & \textbf{5.54\%}  & 0.21\%  & 1.71\%  & 0.01\%  & 2.10\%  & 1.77\%  & 11.3\% \\
Emergence     & \textbf{0.14\%}  & 0.15\%  & 0.16\%  & 0.04\%  & 0.33\%  & 0.13\%  & 1.0 \% \\
Fragmentation & \textbf{1.42\%}  & 0.27\%  & 20.6\%  & 0.07\%  & 12.7\%  & 5.04\%  & 40.1\% \\
Complex       & \textbf{0.00\%}  & 0.01\%  & 0.03\%  & 0.01\%  & 0.07\%  & 0.02\%  & 0.1 \% \\
Error         & \textbf{1.52\%}  & 0.37\%  & 13.6\%  & 0.08\%  & 13.5\%  & 7.01\%  & 36.1\% \\
Survival      & \textbf{1.37\%}  & 0.21\%  & 4.89\%  & 0.03\%  & 3.95\%  & 0.96\%  & 11.4\% \\
\tableline
Total         & \textbf{10.0\%}  & 1.2\%  & 41.0\%  & 0.23\%  & 32.6\%  & 14.9\%  & N=18297\\
\tableline 
\label{tab:N-stats}
\end{tabular}
\end{center}
\end{table*}

\subsection{Disappearance Events}
\label{sub:DisappearanceEvents}
\begin{figure*}
\includegraphics[height=0.5\textwidth]{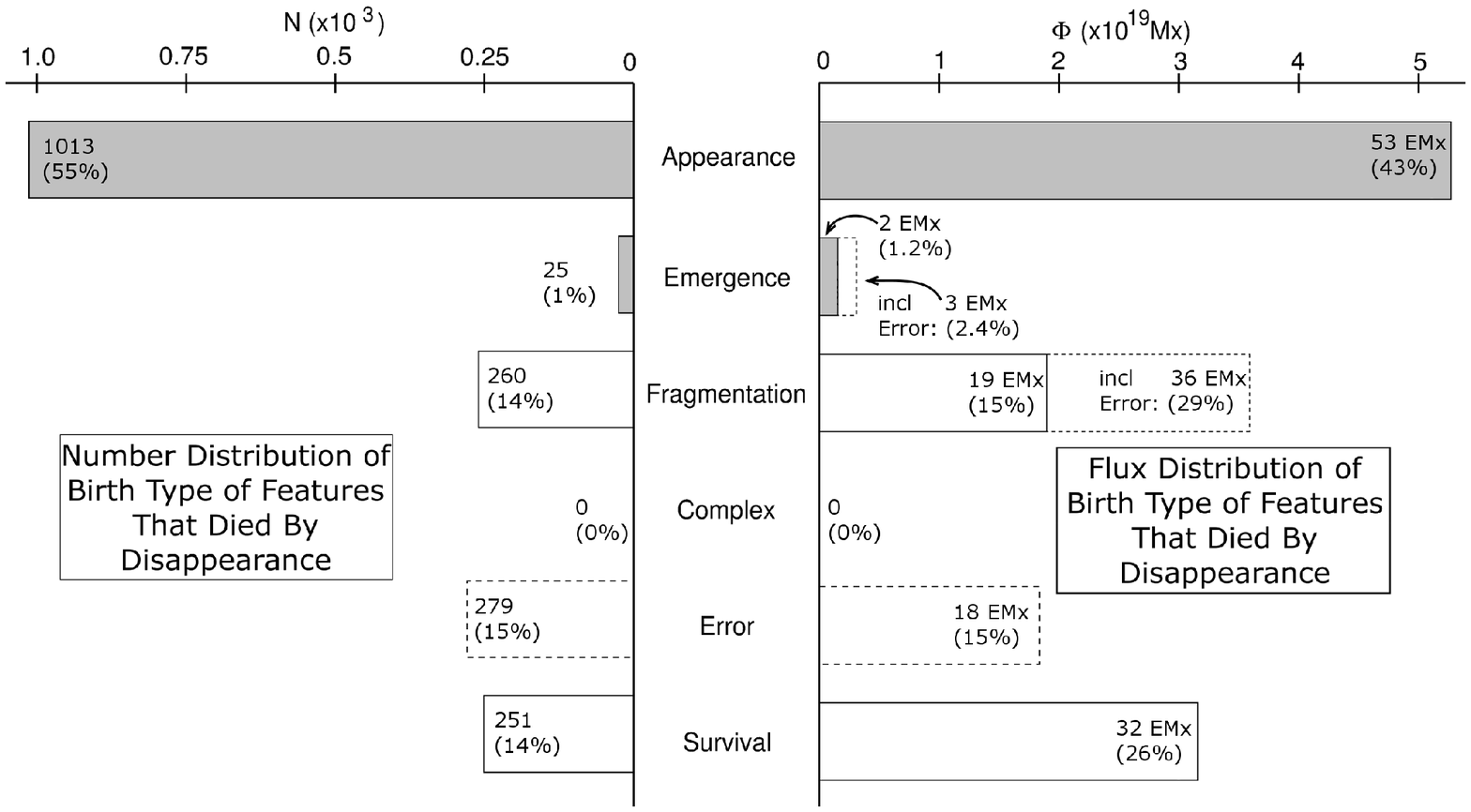}
\caption{
Histograms showing the (left) number and (right) flux distributions of
birth type for the 1828 features that died by Disappearance. These are
in the same format as Figure~\ref{fig:N-Number} except the categories
are by birth instead of death. The flux from the Error birth types
have been distributed across the Emergence and Fragmentation events as
was done for the Cancellation and Merger events in
Figure~\ref{fig:N-Number}, and is again shown by the dashed extensions
to those two bars.
}
\label{fig:disappear-births-histo}
\end{figure*}

As shown in Figure~\ref{fig:N-Number}, Disappearances are responsible
for the removal of the vast majority (83\%) of the flux in the
features identified by SWAMIS. We now investigate the means by which
those features that died by Disappearance (bold font in
Table~\ref{tab:N-stats}) were born.
Figure~\ref{fig:disappear-births-histo} shows the distribution by
(left) number and (right) maximum flux of the birth process of each of
the 1828 features that died by Disappearance. None of the 42 features
with Complex births died by Disappearance, and we do not consider the
Survival and Fragmentation events for reasons given above. The Error
events have been distributed amongst the Emergence and Fragmentation
events in the same way as was done for the Cancellation and Merger
events before.

By far the most common birth process, in both number
and flux, for the Disappearance events is Appearance. Following the
results in \citetalias{Lamb2008} and \citetalias{Lamb2010} these are
features that were born by the coalescence or convergence of smaller
(many smaller than the spatial resolution of NFI) concentrations into
a larger, more magnetically concentrated feature, and died by a
break-up of the larger feature back into smaller concentrations.

\subsection{Disappearing Features that were Not Born by Appearance}
\label{sub:Disapp-no-App}
\begin{figure*}
\includegraphics[height=0.75\textheight]{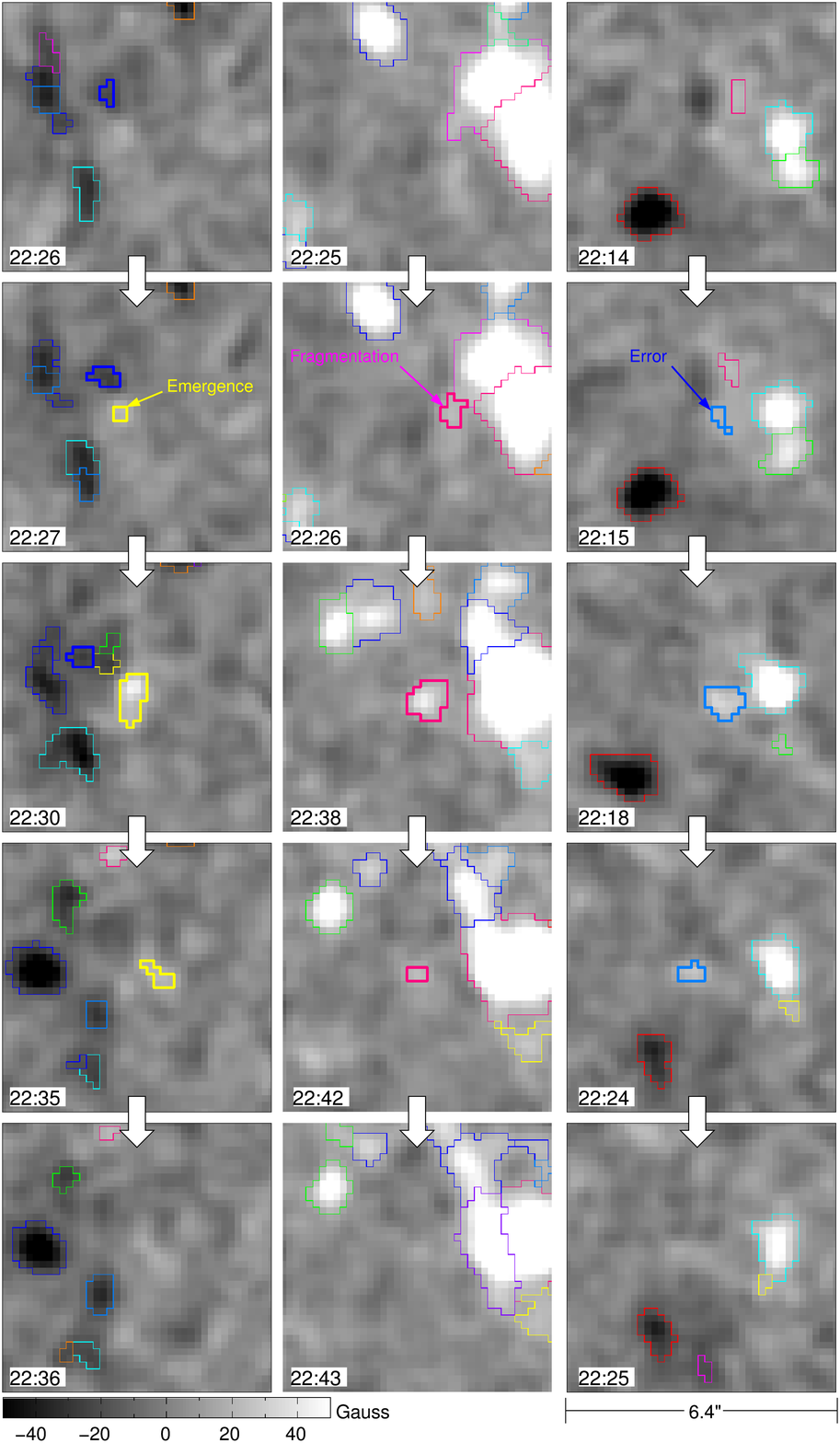}
\caption{\emph{Hinode}/SOT/NFI images of three examples of features that
  died by Disappearance but were born by events other than
  Appearance. The birth events, indicated with the heavy colored
  boundaries, are Left: Emergence; Middle: Fragmentation; Right:
  Error.  The $6.4''\times6.4''$ spatial scale and the
  $\pm50~\textrm{G}$ gray scale are the same as in
  Figure~\ref{fig:death-examples}.  Times in UT are at the bottom left
  of each image.}
\label{fig:disappear-births-examp}
\end{figure*}

It is not surprising that most features that died by Disappearance
were also born by Appearance: in that case, the spatial distribution
of the magnetic field in a relatively weakly-magnetized region changes
such that a new feature is detected, then changes again such that the
feature is no longer detected. However, many features that died by
Disappearance (36\% by number) were born by other event types. This is
more surprising because a feature must be born in a more strongly
magnetized region, move away from other features, and then Disappear.
None of the Complex-born features and only 1.4\%{} of the
Emergence-born features died by Disappearance. Roughly the same
proportion ($\sim$14\%) of the Error-, Survival- and
Fragmentation-born features died by
Disappearance. Figure~\ref{fig:disappear-births-examp} shows examples
of features that were born by Emergence (left column), Fragmentation
(middle column), and Error (right column). Notice that in each case,
the feature is born adjacent to other features, moves away from them,
and then dies by Disappearance. This was typical of these types of
events, and reinforces the idea that Disappearance is merely an
extension, to unobservably small scales, of the shredding process that
causes Fragmentation.

We interpret this result as a macrocosm of the dissipation process described
in \citetalias{Lamb2008} \& \citetalias{Lamb2010}.  These smaller fragments break away from
a larger group, migrate away, and eventually themselves disappear. It
seems likely that this process is due to further dissipation of the
isolated feature in the reverse manner as those that are born by
Appearance.

\subsection{Temporal Ensemble Imaging}
\label{sub:TemporalEnsembleImaging}

In order to test the idea that Disappearance events are typically due
to the dispersal of magnetic flux, rather than unresolved
cancellation, we produced ensemble images of 660 Disappearance events
co-located in space and time.  This technique, Event-Selected Ensemble
Imaging (ESEI), is described in \citetalias{Lamb2008} \&
\citetalias{Lamb2010}. It allows us to discriminate between
Disappearances as asymmetric cancellations between a strong, localized
flux concentration and a larger, undetectably weak opposing region,
versus dispersals of existing flux. We find no evidence of significant
amounts of opposite-polarity flux in the neighborhood of these
Disappearances, similar to our previous analysis of Appearances.

To ensure that we fully understood the ensemble images themselves we
produced an event-selected ensemble movie showing how the ensemble
varied in the time steps surrounding the features' deaths.  For each
Disappearance, we extracted a subimage of the magnetogram image in the
time range $t_{death}-5 ... t_{death}+10$ minutes relative to the time
of Disappearance. The center of the field-of-view of each subimage was
initially taken to be the location of the center of flux of the
Disappearing feature \emph{in the last frame the feature was
  visible}. We call this method of producing the ensemble a ``direct
ensemble''.  For the time of death and the 10 minutes afterwards, this
is the best that can be reliably done. But for the times up to 5
minutes before the death, it can lead to unexpected
results. Specifically, since the feature locations are moving through
$x-y-t$ space in our dataset, in the ensemble image Disappearances
seem to be increasing in strength during the 5 minutes leading up to
the feature death, evidenced by an increasing peak and a smaller FWHM,
a counter-intuitive result. This concentration over time is due to the
uncorrelated (and uncorrected) motions of all the individual features
in the ensemble.

Next, we formed a ``motion-corrected ensemble'' by choosing the center
of the field of view of each subimage to coincide with the measured
center of flux of each feature in the pre-death images. This
eliminated the apparent concentration of the ensemble leading up to
the death event, better approximating a typical feature's
behavior. The peak of the motion-corrected ensemble greatly exceeds
the lower detection threshold (18~G) in the minutes before the death,
slightly exceeds that threshold at the moment of death, and quickly
drops below the threshold after the death.

Since the location of the feature's center of flux is not measured
after the feature has died, and kinematic estimates are unreliable, it
is impossible to apply this same technique to the time after the
feature death has occurred. However, by examining the difference
between the two cases for the time before the death, we can estimate
how much of the post-death spreading of the ensemble is due to
translational motion of the newly-undetected flux, and how much is due
to true dispersal of the flux.

\begin{figure*}
\includegraphics[width=0.70\columnwidth]{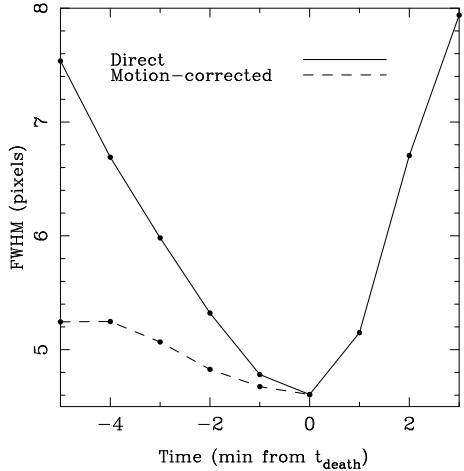}  
\caption{Time evolution of the ensemble FWHM for the cases of a direct
  ensemble (solid line) and a motion-corrected
  ensemble (dashed line), determined by fitting a Gaussian to the core
  of the ensemble. After $t_{death}+3$~min the fit becomes unreliable
  and is not shown.}
\label{fig:fwhm_evol} 
\end{figure*}

In the 5 minutes prior to death, the FWHM of the direct ensemble
decreases approximately linearly from 7.6~pixels at $t_{death}-5$
minutes to 4.6~pixels at $t_{death}$ (Figure~\ref{fig:fwhm_evol}). The
FWHM of the motion-corrected ensemble is smaller, and decreases from
5.25~pixels to 4.6~pixels at $t_{death}$. By definition, the ensembles
are the same at $t_{death}$.  The difference between the two FWHMs is
roughly linear in time over the range $t_{death}-4 \le t \le t_{death}-1$ and
has a slope of 0.44 pixels per frame, or about $0.85\textrm{ km
  s}^{-1}$.

After $t_{death}$ only the uncorrected FWHM is available. We do not
attempt to extrapolate the motion of the feature in the few frames
post-death, since this is unreliable: feature motion in adjacent
frames is not correlated. The uncorrected FWHM increases from the
minimum at $t_{death}$ approximately linearly with time, though the
slope is 50--75\% greater than the pre-death uncorrected slope.

The behavior of the motion-corrected ensemble FWHM is not consistent
with any kind of diffusion of the ensemble magnetic field
distribution.  In the case of normal or anomalous diffusion, the FWHM
would increase proportionally to some (positive) power of the elapsed
time \citep{Abramenko2011}, whereas in our case the FWHM decreases
with time.  In order to understand this, we note that 1) only in the
ensemble average of Disappearance events (and not in the individual
events themselves) is the magnetic field roughly Gaussian-distributed
on the surface; 2) the ensemble FWHM is not the same as the mean of
the individual squared displacements; 3) the field being concentrated
in intergranular lanes means that any diffusion would not be fully
two-dimensional; 4) our motion-correction is based on the
flux-weighted centroid of only those pixels above the low detection
threshold (18~G); 5) the formation of a (meso-)granule around the time
of $t_{death}$ would cause the proper motion of the features to
increase in the minutes after $t_{death}$, and could result in the
increase of the uncorrected FWHM slope at positive times in
Fig.~\ref{fig:fwhm_evol}.

Finally, we note an unanticipated behavior of the Disappearance
ensemble images, which is that the flux that can be confidently
assigned to the Disappearing features (and not to the background) is
not conserved either before or after $t_{death}$. This is not a
by-product of the process that produces the ensemble, since the flux
in individual Disappearance events also cannot be fully accounted
for. Recognizing that our detection thresholds may be high, we
produced images of some Disappearance events with a stretched-out gray
scale and manually enclosed what we believed to be the largest
possible extent of the Disappearing feature. Regardless, the total
enclosed flux always decreased (by 30--50\%) in the minutes after the
feature's death.

We consider likely explanations for this behavior to include 
1) dispersal of the feature's flux via horizontal advection of the
field lines on very small scales, in a process that is completely
analogous to the Fragmentation/Disappearance example in
Figure~\ref{fig:disappear-births-examp};
2) non-linearity in the magnetograph detector, so that a decrease in
the average line-of-sight field in the area of the photosphere
subtended by a pixel does not result in a proportional decrease in the
reported V/I signal and the observed magnetic field strength---such an
effect is expected with the single line-wing magnetograms produced by
NFI, and could account for the loss.

\subsection{Feature Lifetimes \& Partial Lifetimes}
\label{sub:PartialLifetimes}

\begin{figure*}
\includegraphics[width=0.4\textwidth]{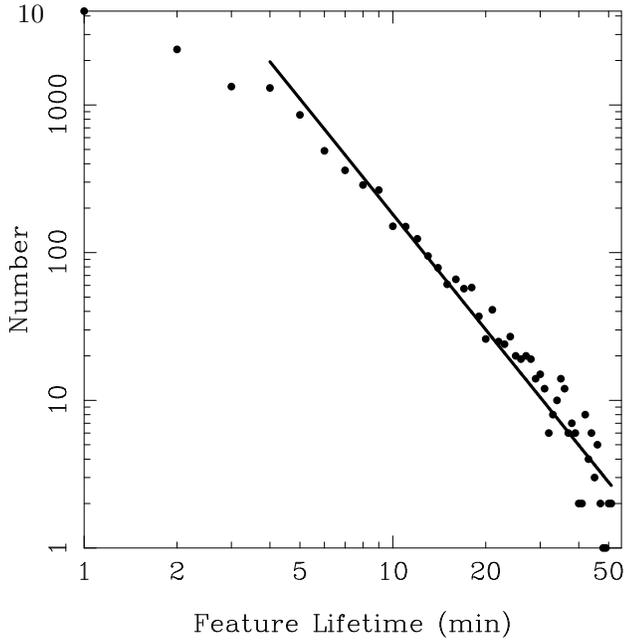}
\caption{Distribution of feature lifetimes on a log-log plot, for only
  those features whose entire lifetime is observed in the dataset. The
  best-fit straight line has a slope $= -2.6$ for lifetimes $\ge 4$
  min. The lower cutoff was chosen because a minimum feature lifetime
  of 4 frames was chosen in the tracking (\S~\ref{sub:Data}); some
  features have a lifetime smaller than this cutoff because we do not
  filter features that are immediately adjacent to other features. The
  linear fit on a log-log plot suggests a power-law lifetime
  distribution, in contrast with the exponential lifetime distribution
  found by \citet{Zhou2010} using the same NFI data. See
  \S~\ref{sub:FeatureHistoryLifetimes} for more discussion.}
\label{fig:LifetimeDist}
\end{figure*}

Figure~\ref{fig:LifetimeDist} shows the distribution of feature
lifetimes for those features that both were born and died during the
observation. The line of best fit for lifetimes $\ge 4$ min on the
log-log plot has a slope of -2.6. The mean (first moment) of a
power-law distribution is finite for slopes $<-2$, and in this case
the best-fit mean time is 10.7~min. It is clear that the lifetime
distribution is not exponential, a point to which we return in
\S~\ref{sub:FeatureHistoryLifetimes}.

Since the total feature death rate is the sum of the rates of death by
all causes, its mathematical reciprocal -- the feature lifetime -- is
the harmonic sum of the ``partial lifetimes'' formed by taking the
reciprocals of the various death processes.  The total death rate by
Cancellation in our observed patch of Sun was
$3.5\times10^{19}\textrm{ Mx h}^{-1}$. Dividing by the time-averaged
total observed unsigned flux of $7.8\times10^{20}\textrm{ Mx}$ yields
a normalized Cancellation rate of $0.045~\textrm{h}^{-1}$, or a
partial lifetime via Cancellation of $22~\textrm{h}$. This is
$\approx3$ times longer than the total lifetime of 8~h estimated for
solar minimum by \citet{Hagenaar2003} based on MDI data, which
included all types of feature death---even Disappearance. The larger
lifetime is especially surprising considering the general trend toward
shorter turnover times as resolution increases, and the lower
resolution of MDI data compared to Hinode. This slower turnover rate
omits consideration of cancellation at smaller scales, which will be
considered in more detail in a subsequent paper. We discuss the
implications of this $22~\textrm{h}$ partial lifetime due to
Cancellation at the end of \S~\ref{sec:Discussion}.

\subsection{Summary of Main Results}
\label{sub:ResultsSummary}

We summarize our main results here, and discuss their implications in
the next section:
\begin{enumerate}
\item Disappearance events account for 10\% of the observed feature
  deaths in our dataset but 83\% of the flux lost to detection
  (\S~\ref{sub:DisappearanceEvents}), although the primary mechanism
  for Disappearance does not eliminate flux from the Sun;
\item Of those Disappearances, more than 50\% were born by Appearance,
  suggesting a large amount of flux constantly being moved into and
  out of our range of detectability (\S~\ref{sub:DisappearanceEvents});
\item We find no evidence that the Disappearance events are largely
  due to undetected cancellation, which agrees with our previous work on
  Appearances (\S~\ref{sub:TemporalEnsembleImaging});
\item The FWHM of the motion-corrected ensemble of Appearances
  decreases with time up to the moment of death, suggesting that
  normal planar diffusion is not an adequate description of this type
  of death event (\S~\ref{sub:TemporalEnsembleImaging});
\item The partial feature lifetime attributed to those features that
  died by Cancellation is 22~h, a factor of 3 slower than previously
  published quiet-Sun turnover times of 8~h
  (\S~\ref{sub:PartialLifetimes}).
\end{enumerate}

\section{Discussion}
\label{sec:Discussion}

\subsection{The Reality of the Disappearance Events}
\label{sub:RealityDisappearances}
A skeptical reader may wonder whether the fact that over half of the
Disappearing features were born by Appearance suggests that these
features could be more confidently attributed to a noise source and
not to physical evolution of the magnetic field.  Photon noise could
hide an opposing pole around the Disappearing features, but as in
\citetalias{Lamb2008} and \citetalias{Lamb2010} the ensemble imaging
(\S~\ref{sub:TemporalEnsembleImaging}) directly addresses that for all
reasonable values of the magnetic pole asymmetry. Other sources of
noise have been mitigated in the data preprocessing and tracking
parameter selection. These noise sources could include, for example, a
temporary and spatially isolated change in the magnetogram noise level
(since the tracking code's detection threshold values do not change
over the course of the dataset), a change in the solar vertical
velocity (since imaging magnetographs are sensitive to surface
velocity fluctuations) or some other cause. Any or all of these could
result in a small locus of pixels exceeding the detection threshold
for a short period of time. Such an occurrence would result in the
Appearance and subsequent Disappearance of a large number of features.

Our data preprocessing, feature filtering criteria, and event
selection criteria were carefully chosen to take the above noise
sources into account. For example, the FFT motion filter used in
preprocessing was finely tuned to reduce noise and reject p-modes. The
minimum feature lifetime was chosen to further reduce the effect of
any photospheric line-of-sight velocity signal leaking into the
magnetogram resulting in spurious feature detection.

Additionally, in post-processing we find no evidence that the
Appearing-Disappearing features are anything other than real. First,
and probably most important, a visual inspection of a movie showing
the time and location of Appearances \& Disappearances reveals no
spatio-temporal clusters of these events. Second, the lifetime
distribution of Appearance events is approximately the same as for
other birth types \citepalias{Lamb2008}, and in the present work 18\%
of the Appearing/Disappearing features have lifetimes $\ge$ 10~min,
twice the period of p-modes which would be the most likely source of
such a surface velocity change. Third, we would have likely seen a
similar effect in the MDI Appearances work of \citetalias{Lamb2008}
but did not, and later confirmed using NFI that the MDI Appearances
were real, thereby validating the \citetalias{Lamb2008} work and the
identical technique used in the present paper.  Therefore we believe
there is sufficient evidence that the Appearing-Disappearing features
seen in this work also can be confidently attributed to true physical
evolution of the photospheric magnetic field.

\subsection{Feature History \& Lifetimes}
\label{sub:FeatureHistoryLifetimes}
Other aspects of the feature history and lifetimes bear mentioning. First,
we note that none of the features that died by Disappearance were of
the Complex birth type. This is understandable when considering the
definition of a Complex birth: there must be at least two nearby
features, a like-polarity feature that satisfies the Fragmentation
criterion and an opposite-polarity feature that satisfies the
Emergence criterion. In order for a feature to be born in such a way,
the local surface density of features must be relatively high. Since
features do not migrate much over the course of their lifetime, the
chance of such a feature dying in the complete absence of other
features is very small.

The same logic applies to the fact that few of the features that die
by Disappearance were born by Emergence, though since there is only one
requirement for an Emergence birth, there are more Emergences than
Complexes, and so by chance some of the features in our dataset have
managed to migrate sufficiently away from their associated birth
feature.  

Finally, we note that the \emph{Hinode}/SOT/NFI dataset used here is
the same as that used in \citetalias{Lamb2010}, and is also the same
as one of the two hour-long NFI datasets used in the by-eye analysis
of \citet{Zhou2010}. By comparing the bottom row of our
Table~\ref{tab:N-stats} or the left panel of Figure~\ref{fig:N-Number}
to the bottom half of their Table~1, it is immediately obvious that
their distribution of death events by type does not agree with our
results. For example, fully \twothirds\ of their deaths were
Disappearances (they used the label F$_{\textrm{situ}}$), compared to
only 10\% of our feature deaths, and 11\% of their deaths were
cancellation, compared to only 1\% of ours. We attribute this
discrepancy to two factors. First, the event definitions used by us
and them are not exactly the same. For example, we have no equivalent
to their F$_{\textrm{frag}}$ because we do not consider a feature to
have died just because another feature was born by Fragmenting off a
small portion of it, whereas they consider this to be the death of one
feature and the birth of two different features. They have no
equivalent to our Error or Complex, which may be due to the ability of
the human brain to categorize complicated edge cases, and to our
attempt to enforce physical flux conservation on events when
possible. Second, they masked out stronger network concentrations and
focused solely on the internetwork magnetic features, whereas we
included all detected features in our feature tracking. The larger
network concentrations exhibit much internal reorganization which
results in many more Merger death events (their F$_{\textrm{coal}}$)
than they observe with the internetwork features. Assuming for a
moment that all of our Mergers and death Error events are in network
concentrations while all other events are in internetwork areas, 38\%
of the remainder of our deaths are Disappearances, which is closer to
their result of 66.6\%. Their masking out of network concentrations is
also likely a main reason why their lifetime distribution does not
agree with ours. They show (in their Figure~4) a lifetime distribution
that is exponential, whereas our feature lifetime distribution is
clearly not an exponential, but is closer to a power-law
(Figure~\ref{fig:LifetimeDist}).  We have previously shown
\citepalias{DeForest2007} that the feature lifetime distribution is
extremely difficult to reconcile between different algorithms (human
or automated) even when operating on the same data with the intent of
reproducing other algorithm's results. We again emphasize that extreme
caution must be employed when drawing scientific conclusions based on
the \emph{distribution} of magnetic feature lifetimes.

\subsection{Concluding Remarks}
\label{sub:ConcludingRemarks}
Our results are consistent with the picture that arises from recent
work by \citet{Parnell2009}, who found that the number of magnetic
features at a given flux scale follows a power law (and therefore has
a scale-invariant distribution) over all observable scales.  At any
particular spatial resolution, the photospheric field is primarily
contained in unobservably small packets of flux that migrate around
the photosphere. Coalescence and shredding of these small packets are
the dominant processes by which the observed visual magnetic features
are created and destroyed.  Far less flux submerges or emerges through
the photosphere on observable scales than moves up or down the range
of available scales, crossing the threshold of observability with any
particular magnetograph.

This constant movement of magnetic flux up and down a range of spatial
scales, into and out of the range of detectability of an instrument
and/or detection algorithm, affects estimates of the ``turnover time''
of photospheric flux, which drives many coronal heating models.
Estimates from feature creation rates \citep[e.g., the 8--19 hr
  reported by ][]{Hagenaar2003} are not necessarily indicative of flux
turnover because of the difference between feature birth and flux
emergence \citepalias{Lamb2008}; and feature average lifetimes are
strongly dependent on the tracking algorithm
\citepalias{DeForest2007}. Depending on the tracking algorithm and the
resulting lifetime distribution, the feature average lifetime may even be
undefined (for a power-law distribution with a slope $\ge -2$).

We have introduced the concept of the partial feature lifetime,
whereby the lifetime due to a particular feature death event type can
be separated from the lifetime distribution as a whole. Traditional
estimates of the flux turnover time are most closely related to the
partial lifetime due to Cancellation, since flux must be introduced
\emph{and removed} from the photosphere in order for it to ``turn
over''. Our partial lifetime due to Cancellation (22~h) is $\approx3$
times slower than previous estimates of the flux turnover time, which
suggests that the shredding and dispersal of flux down to currently
unobservable scales may be more important for determining the lifetime
of magnetic features, than are the more familiar emergence and
cancellation of flux.  The true flux turnover time will depend
strongly on how much magnetic flux is threading the photosphere at
scales currently unobservable to imaging magnetographs, as well as the
manner in which the flux evolves.

The result that most removal of visible flux occurs by shredding it to
smaller scales may make nanoflare heating models simpler to sustain,
both by enabling faster reconnection rates via the small interaction
scales and by supplying several times the visible flux at smaller
scales.  We speculate that there is nothing physically special about
the spatial resolution available to \emph{Hinode} and that the same
scaling law found by \citet{Parnell2009} will proceed to the diffusion
length scale in the photosphere.

\acknowledgements{*}
The authors thank Mandy Hagenaar for insightful early discussions.  DAL was partially supported by NASA grant NNX11AP03G. TAH and CED were partially supported by NASA grant NNX08AJ06G.


\begin{thebibliography}

\bibitem[Abramenko et~al.(2011)]{Abramenko2011} Abramenko, V.~I., 
Carbone, V., Yurchyshyn, V., et al.\ 2011, \apj, 743, 133

\bibitem[Berger et~al.(1998)]{Berger1998} Berger, T.~E., Löfdahl,
  M.~G., Shine, R.~S., \&\ Title, A.~M., 1998, \apj, 495, 973

\bibitem[Cadavid et al.(1999)]{Cadavid1999} Cadavid, A.~C.,
  Lawrence, J.~K., \& Ruzmaikin, A.~A.\ 1999, \apj, 521, 844

\bibitem[DeForest et~al.(2007)]{DeForest2007} DeForest, C.~E., Hagenaar,
H.~J., Lamb, D.~A., Parnell, C.~E., \&\ Welsch, B.~T., 2007,
\apj, 666, 576

\bibitem[De Rosa et~al.(2009)]{DeRosa2009} De Rosa, M.~L., 
Schrijver, C.~J., Barnes, G., et al.\ 2009, \apj, 696, 1780 

\bibitem[Hagenaar(2001)]{Hagenaar2001} Hagenaar, H.~J.\ 2001, \apj,
  555, 448

\bibitem[Hagenaar \&\ Cheung(2009)]{Hagenaar2009} Hagenaar,
  H.~J. \&\ Cheung, M., 2009, ASP Conf. Ser. 415, 167

\bibitem[Hagenaar et~al.(2008)]{Hagenaar2008} Hagenaar, H.~J.,
  De~Rosa, M.~L., \&\ Schrijver, C.~J., 2008, \apj, 678, 541

\bibitem[Hagenaar et~al.(2003)]{Hagenaar2003} Hagenaar, H.~J.,
  Schrijver, C.~J., \& Title, A.~M.\ 2003, \apj, 584, 1107

\bibitem[Hagenaar et~al.(1999)]{Hagenaar1999} Hagenaar, H.~J.,
  Schrijver, C.~J., Title, A.~M., \&\ Shine, R.~A., 1999, \apj, 511,
  932

\bibitem[Harvey \& Harvey(1973)]{HarveyHarvey1973}
Harvey, K.~L. \& Harvey, J.~W., 1973, Solar Phys., 28, 61

\bibitem[Harvey(1993)]{Harvey1993PhDT} Harvey, K.~L.\ 1993, 
Ph.D.~Thesis

\bibitem[Hathaway et~al.(1996)]{Hathaway1996} Hathaway, D.~H., Gilman,
P.~A., Harvey, J.~W., Hill, F., Howard, R.~F., Jones, H.~P., Kasher,
J.~C., Leibacher, J.~W., Pintar, J.~A., \&\ Simon, G.~W., 1996,
Science, 5266, 1306

\bibitem[Iida et~al.(2012)]{Iida2012} Iida, Y., Hagenaar, H.~J., 
\& Yokoyama, T.\ 2012, \apj, 752, 149

\bibitem[Krivova \& Solanki(2004)]{Krivova2004} Krivova, N.~A., \&\ Solanki,
S.~K., 2004, Astron. Astrophys. 417, 1125

\bibitem[Lamb et~al.(2008)]{Lamb2008} Lamb, D.~A., DeForest, C.~E.,
Hagenaar, H.~J., Parnell, C.~E., \&\ Welsch, B.~T., 2008, Astrophys.
J., 674, 520

\bibitem[Lamb et~al.(2010)]{Lamb2010} Lamb, D.~A., DeForest, C.~E.,
Hagenaar, H.~J., Parnell, C.~E., \&\ Welsch, B.~T., 2010, Astrophys.
J., 720, 1405

\bibitem[Leighton(1964)]{Leighton1964} Leighton, R.~B., 1964,
\apj, 140, 1547

\bibitem[Longcope \& Kankelborg(1999)]{Longcope1999} Longcope, D.~W.,
\&\ Kankelborg, C.~C., 1999, \apj, 524, 483

\bibitem[L{\'o}pez Fuentes et~al.(2006)]{LopezFuentes2006} L{\'o}pez 
Fuentes, M.~C., Klimchuk, J.~A., \& D{\'e}moulin, P.\ 2006, \apj, 639, 459

\bibitem[Muller et~al.(2000)]{Muller2000} Muller, R., Dollfus, A.,
Montague, M., Moity, J., \&\ Vigneau. J., 2000, Astron. Astrophys.,
339, 373

\bibitem[Kosugi et al.(2007)]{Kosugi2007} Kosugi, T., Matsuzaki, 
K., Sakao, T., et al.\ 2007, \solphys, 243, 3

\bibitem[Livi et al.(1985)]{Livi1985} Livi, S.~H.~B., Wang, J., 
\& Martin, S.~F.\ 1985, Australian Journal of Physics, 38, 855

\bibitem[Parker(1988)]{Parker1988} Parker, E.~N.\ 1988, \apj, 330, 
474

\bibitem[Parnel \& Jupp(2000)]{ParnellJupp2000} Parnell, C.~E. \&\ Jupp,
P.~E., 2000, \apj 529, 554

\bibitem[Parnell(2001)]{Parnell2001} Parnell, C.~E., 2001, Solar
Phys., 200, 23

\bibitem[Parnell et~al.(2009)]{Parnell2009} Parnell, C.~E., DeForest, C.~E.,
Hagenaar, H.~J., Johnston, B.~A., Lamb, D.~A., \&\ Welsch, B.~T. 2009, \apj 698, 75

\bibitem[Pevtsov \& Acton(2001)]{Pevtsov2001} Pevtsov, A.~A., \&\ Acton,
L.~W., 2001, \apj, 554, 416

\bibitem[Pietarila Graham et~al.(2009)]{PietarilaGraham2009} Pietarila 
Graham, J., Danilovic, S., \& Sch{\"u}ssler, M.\ 2009, \apj, 693, 1728

\bibitem[Priest et~al.(1994)]{Priest1994} Priest, E.~R., Parnell,
C.~E., \&\ Martin, S.~F., 1994, \apj, 427, 459

\bibitem[Scherrer et al.(1995)]{Scherrer1995} Scherrer, P.~H., 
Bogart, R.~S., Bush, R.~I., et al.\ 1995, \solphys, 162, 129

\bibitem[Schrijver et~al.(1997)]{Schrijver1997} Schrijver, C.~J.,
  Title, A.~M., van~Ballegooijen, A.~A., Hagenaar, H.~J., and Shine,
  R.~A., 1997, \apj 487, 424

\bibitem[Schrijver \& Zwaan(2000)]{Schrijver2000} Schrijver, C.~J.,
\&\ Zwaan, C., 2000, Solar and Stellar Magnetic Activity, Cambridge
Uni. Press

\bibitem[Simon et~al.(1995)]{Simon1995} Simon, G.~W., Title, A.~M.,
\&\ Weiss, N.~O., 1995, \apj, 442, 886

\bibitem[Simon et~al.(2001)]{Simon2001} Simon, G.~W., Title, A.~M., \&\ Weiss, N.~O.,
2001, \apj, 561, 427

\bibitem[Smithson(1973)]{Smithson1973} Smithson, R.~C., 1973, Solar
Phys., 29, 365

\bibitem[Trujillo Bueno et~al.(2004)]{TrujilloBueno2004} Trujillo Bueno, 
J., Shchukina, N., \& Asensio Ramos, A.\ 2004, \nat, 430, 326

\bibitem[Tsuneta et~al.(2008)]{Tsuneta2008} Tsuneta, S. et~al. 2008: Sol. Phys. 249, 167

\bibitem[Wang et~al.(1988)]{Wang1988} Wang, J., Shi, Z., Martin,
S.~F., \&\ Livi, S.~H.~B., 1988, Vistas Astron., 31, 79

\bibitem[Welsch et~al.(2007)]{Welsch2007shootout} Welsch, B.~T., Abbett, 
W.~P., De Rosa, M.~L., et al.\ 2007, \apj, 670, 1434 

\bibitem[Welsch et al.(2013)]{Welsch2013} Welsch, B.~T., Fisher, 
G.~H., \& Sun, X.\ 2013, \apj, 765, 98 

\bibitem[Zhou et~al.(2010)]{Zhou2010} Zhou, G.~P., Wang, J.~X.,
\&\ Jin, C.~L., 2010, Solar Phys., 267, 63

\end{thebibliography}
\end{document}